\documentclass[twocolumn, tighten, twocolappendix]{aastex63}
\accepted{November 10, 2020}
\submitjournal{ApJ}
\shorttitle{MAPPIES I - Description and First Results}
\shortauthors{Dreyer and B{\"o}ttcher}
\graphicspath{{./}{figures/}}
\usepackage{amsmath}
\begin{document}

\title{Monte-Carlo Applications for Partially Polarised Inverse External-Compton Scattering (MAPPIES) - I. Description of the code and First Results}
\correspondingauthor{Lent{\'e} Dreyer}
\email{lentedreyer@gmail.com}

\author[0000-0002-4971-3672]{Lent{\'e} Dreyer}
\affiliation{Centre of Space Research, North-West University, Potchefstroom 2531, South Africa}
\author{Markus B{\"o}ttcher}
\affiliation{Centre of Space Research, North-West University, Potchefstroom 2531, South Africa}

\begin{abstract}

The radiation mechanisms responsible for the multiwavelength emission from relativistic jet sources are poorly understood. The modelling of the spectral energy distributions (SEDs) and light curves alone is not adequate to distinguish between existing models. Polarisation in the $X$-ray and $\gamma$-ray regime of these sources may provide new and unique information about the jet physics and radiation mechanisms. Several upcoming projects will be able to deliver polarimetric measurements of the brightest $X$-ray sources, including active galactic nuclei (AGN) jets and $\gamma$-ray bursts (GRBs). This article describes the development of a new Monte-Carlo code -- MAPPIES (Monte-Carlo Applications for Partially Polarised Inverse External-Compton Scattering) -- for polarisation-dependent Compton scattering in relativistic jet sources. Generic results for Compton polarisation in the Thomson and Klein-Nishina regimes are presented. 

\end{abstract}
\section{Introduction} \label{sec:INTRO}
The radiation from jetted astrophysical sources (e.g. active galactic nuclei (AGNs) and $\gamma$-ray bursts (GRBs)) is characterised by their spectral energy distribution (SED) which can be modelled in many different ways, whilst being consistent with the spectral shape of the SED (e.g. \cite{Burrows_etal2006, Walcher_etal2011, Bottcher_etal2013}). Additional constraints are therefore required in order to distinguish between models. Relativistic jets are accompanied by the acceleration of particles up to very high energies, as well as the production of secondary, non-thermal radiation \citep{Bottcher_etal2012}. Understanding the particle acceleration, radiation mechanisms, and the magnetic field structure of these jets is among the primary goals in the field of high-energy astrophysics. Polarisation is fundamentally linked to the internal geometry of astrophysical sources, and therefore carries important information about the astrophysical environment in terms of the how the magnetic field is linked into the dynamics and acceleration of the energetic particles (see \cite{Trippe_2014} for a general review). Polarisation of the emission from jet-like astrophysical sources adds two essential parameters -- the polarisation degree (PD) and the polarisation angle (PA) -- to those already derived from spectra and variability.

Synchrotron polarisation of the radio/optical emission from astrophysical jets has been a standard diagnostic to examine the magnetic field (e.g. \cite{Conway_etal1993, Zhang_etal2014, Gabuzda_2018}), since the polarisation measurements combined with the spectra and variability of the emission reveal critical information about the magnetic field structure in the emission region. However, since the radio/optical polarisation may come from regions that do not emit high-energy radiation, an important challenge that remains for high-energy astronomy is measuring the polarisation in the ultraviolet (UV), $X$-ray, and $\gamma$-ray regimes in order to probe the most active jet regions with powerful particle acceleration. High-energy polarisation measurements may provide unambiguous constraints on the geometry and structure of the astrophysical source, for example by constraining the orientation of the accretion-disks with respect to our line of sight (e.g. \cite{Schnittman_Krolik_2010, Schnittman_Krolik_2010b, Laurent_etal2011, Beheshtipour_etal2017}). Compared to the orientation of the relativistic jet, multiwavelength polarisation, therefore, holds vital 
\clearpage
\begin{deluxetable*}{lcll}
\tablenum{1}
\tablecaption{List of recent, upcoming, and future proposed missions to measure the polarisation of the high-energy emission from jet-like astrophysical sources. \label{tab:PolMissions}}
\tablewidth{0pt}
\tablehead{
\colhead{\textbf{Polarimeter}} & \colhead{\textbf{Energy} [keV]} & \colhead{\textbf{References}} & \colhead{\textbf{Science objectives}} 
}
\startdata
\hline
Lightweight Asymmetry and Magnetism  & $0.25$ &  \cite{LAMP_2015} & Blazar jets, and thermal \\
Probe (LAMP) & &  \cite{LAMP_2019} & emission from pulsars \\
& & & \\
Experiment Demonstration of a Soft  & $0.2 - 0.8$ & \cite{REDSoX_2017} & AGNs, binaries, and \\
$X$-ray Polarimeter (REDSoX) & & \cite{REDSoX_2019} & isolated pulsars \\
& & & \\
The $X$-ray Polarization Probe (XPP) & $0.2 - 60$ & \cite{XPP_2019} & Black holes, neutron stars, \\
&  & \cite{XPP_2019} & magnetars, and AGNs\\
& & & \\
Enhanced $X$-ray Timing and Polarimetry  & $2.0 - 10$ & \cite{eXTP_2016} & Black holes, neutron stars, \\
Mission (eXTP) & & \cite{eXTP_2019} & and AGNs\\
& & & \\
Imaging $X$-ray Polarimetry Explorer (IXPE) & $2.0 - 8.0$ & \cite{IXPE_2016} & AGNs, black holes, \\
& & \cite{IXPE_2019} & and neutron stars. \\
& & & \\
POLAR and POLAR-2 & $0.05 - 0.5$ & \cite{POLAR_2018}& GRBs \\
& & \cite{POLAR2_2019} & \\
& & & \\
$X$-ray Polarimeter Experiment (POLIX) & $5.0 - 30$ &\cite{POLIX_2016} & AGN jets, black holes, and  \\
&  & & accretion powered pulsars. \\
& & & \\
X Calibur and XL-Calibur & $20 - 40$ & \cite{XLCalibur_2019} & AGN jets, black holes, and \\
& & \cite{XLCalibur_2020} & neutron stars.\\
\enddata
\end{deluxetable*}
\noindent information on the extreme physical processes and morphology of the sources and their jets (see e.g. \cite{Krawczynski_etal2011, Zhang_2017, Krawczynski_etal2019, Liodakis_etal2019, Bottcher_2019, Rani_etal2019} for reviews). 

The general formalism for calculating high-energy polarisation has been well established. Synchrotron radiation of relativistic charged particles in ordered magnetic fields is expected to be both linearly and circularly polarized \citep{Westfold_1959, Rybicki_Lightman_1979}. Compton scattering off relativistic electrons will reduce the degree of polarisation to about half of the seed photon field's polarisation \citep{Bonometto_etal1970}. Since there is no existing technology to measure high-energy circular polarization, and the radiation is treated as a collection of particles, rather than an electromagnetic wave, only linear polarisation is considered. The Klein-Nishina cross section is generally dependent on the polarisation. Polarised photons scatter preferentially in a direction perpendicular to their electrical field vector, and the electric field vectors of the scattered photons tend to align with the seed photons' electric field vectors \citep{Matt_etal1996}. Polarisation can therefore be induced when non-relativistic electrons scatter an anisotropic photon field, even if the seed photons are unpolarised. Compton scattering off relativistic electrons, however, is not expected to induce polarisation since the seed photon field is approximately axisymmetric around the electron momentum in the electron rest frame, making any anisotropy of the photon field irrelevant. In a model where thermal and a power-law tail of non-thermal electrons (in an emission region that moves along the jet with a bulk Lorentz factor $\Gamma_{jet} \geq 10$) scatter an external optical/UV radiation field, the hard $X$-ray/$\gamma$-ray radiation will results from scattering off relativistic electrons and is thus expected to be unpolarised. The UV/soft $X$-ray radiation, on the other hand, results from scattering off non-relativistic thermal electrons and can therefore be highly polarised (e.g. \cite{Schnittman_Krolik_2010b}).  

A formalism for evaluating Compton polarisation in the Thomson regime was developed by \cite{Bonometto_etal1970}, and applied to Synchrotron Self Compton (SSC) emission by \cite{Bonometto_Saggion_1973}. Calculations of Compton polarisation in the Thomson and Klein-Nishina regimes were provided by \cite{Sunyaev_Titarchuk_1984}, following the Monte-Carlo approach. \cite{Krawczynski_2012} and \cite{Beheshtipour_etal2017} also followed the Monte Carlo approach to calculate the Compton polarisation in the Thomson and Klein-Nishina regimes, which included the contribution of non-thermal electrons in the emission region, and verified the analytical results of \cite{Bonometto_etal1970} in the Thomson regime. This article describes the development of a new Monte-Carlo code -- MAPPIES (Monte-Carlo Applications for Partially Polarised Inverse External-Compton Scattering) -- for polarisation dependent Compton scattering in jetted astrophysical sources. 

The potential of using high-energy polarisation as a diagnostic for different radiation mechanisms is briefly discussed in section \ref{sec:HEpol}. In section \ref{sec:MAPPIES}, the description of the MAPPIES code is presented, followed by the results for Compton polarisation in the Thomson and Klein Nishina regimes in section \ref{sec:MSC_Results}, and a summary in section \ref{sec:summary}.
\section{Scientific Potential of High-Energy Polarisation}\label{sec:HEpol}
High-energy polarimetry may play a crucial role in understanding the extreme physics of high-energy radiation, neutrino production, and particle acceleration in jet-like astrophysical sources. Polarisation of $X$-ray/$\gamma$-ray emission has remained largely unexplored, partly due to the difficulty in the detection of high-energy polarisation. However, advancements of new technology lead to several projects that may be able to deliver polarimetric measurements of high-energy emission from the brightest $X$-ray sources, with estimates of a minimum detectable degree of polarisation (MDP) down to $10\%$ for moderately bright sources \citep{McConnell_Ryan_2004}. Examples of recent, upcoming, and proposed missions to measure the polarisation in the high-energy regime of the emission from jet-like astrophysical sources is listed in Table \ref{tab:PolMissions}. It is thus timely to consider the model predictions for different models of various astrophysical sources. In this section, a short overview of the model predictions for the radiation mechanisms of high-energy emission from the AGN jets and GRBs is given.  
  
\subsection{High-energy emission from AGNs}
Jet dominated AGNs, which are among the most powerful high-energy emitters in the Universe, harbor supermassive black holes (SMBHs) at their central engines. In about $10\%$ of AGNs, mass accretion onto this SMBH is accompanied by the production of relativistic jets, whose bulk energy is converted into kinetic energy of electrons, multiwavelength radiation, and possibly particle emission of ions and neutrinos. Blazars, in which one of the jets is closely aligned to our line of sight \citep{Urry_Padovani_1995}, are the most numerous class of extragalactic $\gamma$-ray sources detected (e.g. \cite{Aharonian_etal2009, Ackermann_etal2016}). The SEDs of blazars are dominated by non-thermal radiation across the entire electromagnetic spectrum. The radio through optical/UV emission is well explained by synchrotron emission from relativistic electrons in the jet, which is consistent with moderate PDs (up to PD $\sim(3 - 40)\%$) in the optical \citep{Bottcher_etal2012, Zhang_etal2014}. For high-energy-synchrotron-peaked (HSP) blazars, the synchrotron emission extends well into the $X$-ray regime, and may thus be confirmed by $X$-ray polarisation (e.g. \cite{Krawczynski_etal2011}).

The origin of the high-energy ($X$-ray/$\gamma$-ray) component in the SEDs of blazars is still unclear with two viable models, both consistent with the shape of the SED: The first \textit{leptonic} model argues that the high-energy component is due to Compton scattering off the same electrons that emit the radio to UV/$X$-ray radiation. The seed photons can then either be synchrotron photons or infrared (IR)/optical/UV photons external to the jet (possibly from the accretion disk, the broad line region (BLR), or a dusty torus). In the second \textit{hadronic} model, non-thermal protons radiate via proton synchrotron emission and interact with low-energy photons via photo-pair and photo-pion interactions leading, in most cases, to synchrotron-supported pair cascades developing in the emission region (for a general review on the features of these models, see \cite{Bottcher_2010}). High-energy polarisation provides an excellent diagnostic to distinguish between leptonic and hadronic emission scenarios, since hadronic models intrinsically predict higher degrees of $X$-ray and $\gamma$-ray polarisation than leptonic models \citep{Zhang_Bottcher_2013, Paliya_etal2018}. 

The production of high-energy neutrinos provides evidence for hadronic interactions (e.g. \cite{Atoya_Dermer_2001, Dermer_Menon_2010, Tavecchio_Ghisellini_2015}). If blazars accelerate enough high-energy protons, the protons may interact with the local blazar radiation field and produce charged pions which decay and emit neutrinos. The recent IceCube-170922A neutrino event, which was reported to coincide with the blazar TXS 0506+056 in flaring state \citep{IceCubeCollaboration_2018}, indicates that hadronic processes may operate in a blazar jet. Many models have been put forward for an explanation of the corresponding SED of TXS 0506+056 during the neutrino alert (e.g. \cite{Ansoldi_etal2018, Keivani_etal2018, Murase_2018, Padovani_etal2018, Cerruti_etal2019, Reimer_etal2019, Gao_etal2019}), which can be categorised into two groups: The first is a leptonic scenario where inverse-Compton dominates the high-energy emission, with a subdominant hadronic component which produces the neutrinos as well as a considerable amount of $X$-rays through synchrotron emission from hadronically induced cascades. The second is a hadronic scenario where the $X$-ray emission consists of both proton synchrotron and cascading synchrotron, while the $\gamma$-ray emission is dominated by proton synchrotron. $X$-ray polarisation can probe the secondary pair synchrotron contribution complementary to the neutrino detection, while $\gamma$-ray polarisation can be used to distinguish between the inverse-Compton and proton synchrotron scenarios \citep{Rani_etal2019}. For instance, \cite{Zhang_etal2019} found that the proton synchrotron (hadronic) scenarios generally predict higher PDs across the high-energy component than the inverse-Compton (leptonic) dominated scenarios. 

The SEDs of some blazars contain an excess in the UV and/or soft $X$-ray regime (e.g. \cite{Masnou_etal1992, Grandi_etal1997, Haardt_etal1998, Pian_etal1999, Raiteri_etal2005, Palma_etal2011, Ackermann_etal2012, Paliya_etal2015, Pal_etal2020}) called the big blue bump (BBB). Various models for the emission responsible for the BBB have been proposed, which include: (1) Thermal emission from the accretion disk that feeds the SMBH (e.g. \cite{Pian_etal1999, Blaes_etal2001, Paliya_etal2015, Pal_etal2020}), (2) a higher than Galactic dust-to-gas ratio towards the source (e.g. \cite{Ravasio_etal2003}), (3) a distinct synchrotron component from a different region in a multi-zone construction (e.g. \cite{Paltani_etal1998, Raiteri_etal2006, Ostorero_etal2004, Roustazadeh_Botther_2012}), (4) the detection of a Klein-Nishina effect on the synchrotron spectrum (e.g. \cite{Ravasio_etal2003, Moderski_etal2005}), and (5) bulk Compton emission (e.g. \cite{Sikora_etal1994, Sikora_etal1997, Blazejowski_etal2000, Ackermann_etal2012, Baring_etal2017}). The polarisation of the UV/$X$-ray emission from blazars may yield significant insights in order to distinguish between different BBB emission scenarios. For instance, a BBB due to (unpolarised) thermal emission from an accretion-disk predicts that the polarisation will decrease with increasing frequency throughout the optical/UV regime, while \cite{Roustazadeh_Botther_2012} predicted that a BBB due to cascade synchrotron emission would result in PDs that show a weak dependence on the frequency over the optical/UV regime. If the BBB arises from the bulk Compton feature, the thermal Comptonisation process should lead to significant polarisation of the Compton emission from the UV/$X$-ray excess in the SED \citep{Baring_etal2017}. 
\subsection{Gamma-ray burst prompt emission}
GRBs are the strongest explosions in the Universe, separated into two phases: The initial burst of $\gamma$-rays (i.e. the \textit{prompt} emission) that can last from a fraction of a second to hundreds of seconds, and a longer lasting (from days to weeks) \textit{afterglow} emission. There are at least two classes of GRBs depending on the duration of the the prompt emission phase, believed to be associated with the formation of two oppositely directed ultra-relativistic jets \citep{McConnell_etal2019}: Short GRBs ($\leq 2$ seconds), which are believed to be associated with the merger of compact star binaries, and long GRBs ($> 2$ seconds), which are believed to be associated with the death of a massive star. The afterglow has been well studied across the entire electromagnetic spectrum, contributing in our understanding of the later stages of GRB jets (e.g. \cite{Sari_1997, Bottcher_Dermer_1998, Pian_etal1999, Piran_Granot2001, Granot_2008, Racusin_etal2011}). 

However, given the erratic nature of GRBs, understanding the physics of the early phase of the jet propagation is more challenging since it depends on the short lived high-energy prompt emission. The observed spectra of GRB prompt emission are often well described by a \textit{Band} function \citep{Band_etal1993}, which consists of a broken power-law with a smooth break at a characteristic \textit{peak} energy (i.e. the peak of the spectrum when plotted in terms of the energy output per decade). Many diverse models for the emission mechanism that can explain the Band like non-thermal prompt emission spectra have been proposed (see e.g. \cite{Baring_Braby_2004, Peer_2015}), which include: (1) Optically thin synchrotron radiation with either random or ordered magnetic fields, (2) SSC emission, (3) Compton drag models, and (4) photospheric models. The models show that an integrated understanding of the geometry and physical processes close to the central engine may only be accomplished through high-energy polarimetry, since it depends on the emission processes involved that produce the prompt emission (see e.g. \cite{McConnell_etal2019, Gill_etal2020} for recent reviews). 

Emission due to the SSC process can be moderately polarised, with maximum PDs $\sim 24\%$ for a simplified model \citep{Chang_Lin_2014}. However, SSC has been disfavored as a plausible emission mechanism by GRB energetics (see e.g. \cite{Baring_Braby_2004, Piran_etal2009}). The predicted linear polarisation for photospheric models is relatively low, although the polarisation can be as high as PD $\sim 40 \%$ depending on the line of he sight \citep{Lundman_etal2018}. Models that argue for synchrotron radiation and/or inverse-Compton scattering between softer photons and relativistic electrons (i.e. Compton drag) predict high PDs \citep{Lyutikov_etal2003, Gill_etal2020}. Detection of highly polarised prompt GRB emission would thus support both synchrotron and Compton drag models. However, \cite{Toma_etal2009} showed that a statistical study of a sample of GRBs could then be used to differentiate between the models that either invoke ordered or random magnetic fields. 
\clearpage
\begin{deluxetable*}{ll}
\tablenum{2}
\tablecaption{The input parameters of the MAPPIES code for different seed photon and electron energy distributions. \label{tab:MAPPIES_param}}
\tablewidth{0pt}
\tablehead{\colhead{\textbf{Model description}} & \colhead{\textbf{Parameter description}}}
\startdata
\hline
{Emission region} & Lorentz factor of the jet, $\Gamma_{jet}$ \\
{} & Redshift of the source, $z$ \\
\hline
\hline
\multicolumn{2}{c}{\textbf{Seed photon distributions}}\\
\hline
{(1) Isotropic, single-temperature } & Number of seed photons considered, $N_{phot}$ \\
{black body distribution.} & Temperature of the seed photons in eV, $kT_{rad}$ \\
\hline
{} & Number of seed photons considered, $N_{phot}$ \\
{(2) Multi-temperature} & Black hole mass in g, $M_{BH}$ \\
{accretion-disk} & Inner disk radius of the accretion-disk in cm, $R^{in}_D$ \\
{spectrum.} & Outer disk radius of the accretion-disk in cm, $R^{out}_D$ \\
{} & Distance between the central black hole and the emission region in cm, $h$ \\
{} & Accretion-disk luminosity in $\mathrm{~erg \cdot s^{-1}}$, $L_D$ \\
\hline
\hline
\multicolumn{2}{c}{\textbf{Electron energy distributions}}\\
\hline
{(1) Purely thermal (Maxwell) distribution.} & Thermal temperature of electrons in eV, $kT_{e}$ \\
\hline
{(2) Hybrid (Maxwell + power-law)} & Fraction of non-thermal electrons, $f_{nth}$ \\
{distribution.}& Power-law index of the power-law distribution of non-thermal electrons, $p$ \\
{}& Maximum Lorentz factor of non-thermal electrons, $\gamma_{max}$ \\
\hline
{(3) Input distribution} & File of electron distribution, $n(\gamma)$ \\
\enddata
\end{deluxetable*}
\noindent The fraction of the GRBs with significant polarisation is higher for models that have ordered magnetic fields than those with random magnetic fields. The PD measurements of a sample of GRBs which result in a distribution peaking at high PDs would therefore favor synchrotron models with ordered magnetic fields.  

A compact detector for GRB polarisation, POLAR \citep{POLAR_2018}, was designed to produce linear polarisation measurements in the energy range of $\sim 50 - 500$ keV and detected 55 GRBs during 2016 and 2017 \citep{Xiong_etal2017, Kole_2018}. A time integrated analysis for a number of GRBs observed by POLAR was done by \cite{Kole_2018}, which showed low or unpolarised prompt emission in the energy regime of $\sim 30 -750$ keV. The results therefore favored most emission mechanism models, expect synchrotron radiation with ordered magnetic field configurations. Time resolved analysis for a selected sample of the GRBs revealed PDs with a changing in PA. For instance, the sample of GRBs for which intra-pulse time-resolved studies were possible indicated PDs $\sim 30\%$ with an evolving PA. This indicates that low polarisation signals from the time-integrated analysis could be a result of the summation of changing polarisation signals for different epochs. The work of \cite{Kole_etal2020} did not include energy resolved studies which have the potential to test predictions for different models. Various components of the prompt emission can have different distinct PDs, for example, \cite{Lundman_etal2018} predicted significant polarisation in the energy regime of 10s of keV, but also predicted that the PD can be lost at higher energies due to Comptonisation. Energy-dependent polarisation studies can therefore be a powerful diagnostic for different emission models. The code presented in this paper can be used for energy- and angle-dependent studies of Compton polarisation from relativistic jet sources. 
\section{The MAPPIES code}\label{sec:MAPPIES}
\begin{figure*}[ht!]
\plotone{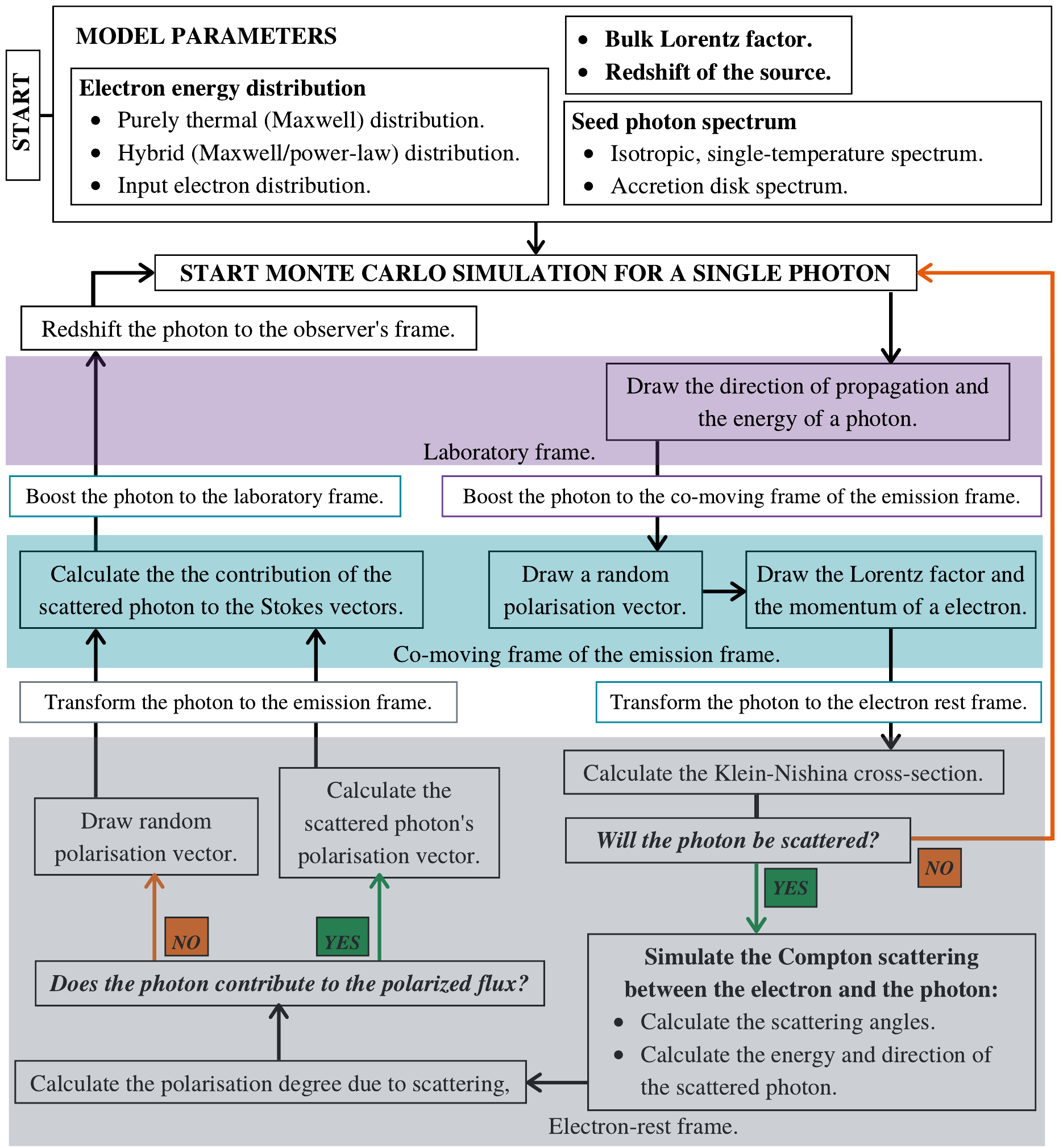}
\caption{A flow diagram for the Monte-Carlo simulation of the MAPPIES code. The input parameters define the characteristics of the emission region (which moves along the jet with a bulk Lorentz factor $\Gamma_{jet}$), the seed photon distribution (drawn from an isotropic, single-temperature black body distribution, or a multi-temperature accretion-disk spectrum), and the electron energy distribution (drawn from a purely thermal distribution, a hybrid distribution, or an input electron spectrum). The Monte-Carlo simulation is shown for a single photon, and will continue for the number of seed photons considered. The photon is transformed between the laboratory frame (shown in the purple shaded area), the emission frame (shown in the blue shaded area), and the electron rest frame (shown in the grey shaded area). After evaluating the full Klein-Nishina cross section in the electron rest frame, the code continues with only the scattered photons. \label{fig:MAPPIES}}
\end{figure*}

In this section, a newly developed Monte-Carlo code (MAPPIES) for polarisation-dependent Compton scattering of radiation fields in relativistic jets of e.g. GRBs and AGNs is presented. A flow diagram of the code is shown in Figure \ref{fig:MAPPIES}: An external radiation field, originating in the laboratory frame (which can be e.g. the rest frame of the AGN, or the rest frame of the GRB progenitor), scatters off an arbitrary (thermal and non-thermal) electron distribution, assumed to be isotropic in the co-moving frame of the emission region that moves along the jet with a relativistic speed (i.e. a bulk Lorentz factor $\Gamma_{jet} \gg 1$). The MAPPIES code is written in the object orientated programming language C++, and performs the radiation transfer simulation by tracking every photon individually (i.e. the \textit{single} photon approach) through the computational domain. The Monte-Carlo single photon approach is very flexible in terms of the Comptonising medium, as well as the directional and energy distributions of the seed photons and electrons. The input parameters (listed in Table \ref{tab:MAPPIES_param}) determine the characteristics of the emission region, the seed photon distribution (discussed in section \ref{subsec:MAPPIES__SeedPhotons}), and the electron energy distribution (discussed in section \ref{subsec:MAPPIES__Electron}). The polarisation signatures are calculated using the Stokes formalism \citep{Stokes_1851}, and the polarisation-dependent Compton scattering of the seed photons is evaluated using Monte-Carlo methods by \cite{Matt_etal1996} (discussed in section \ref{subsec:MAPPIES__Compton}). A comparison to previously published results is given in Appendix \ref{sec:MSC_Compare}.

In what follows, quantities in the laboratory frame and the co-moving frame of the emission region are denoted with subscripts \textit{lab} and \textit{em}, respectively. Quantities in the electron rest frame and the observer's frame are denoted with subscripts \textit{e} and \textit{obs}, respectively, and the scattered photon quantities are denoted with a superscript \textit{sc}.  While following the single photon approach, an additional subscript \textit{i} is used to label the quantities of the current, individual photon. All random numbers are denoted with $\xi$ and are drawn with the \textit{Mersenne Twister} \citep{Mersenne_Twister} between $0$ and $1$, unless specified otherwise. 

\subsection{Seed photon fields}\label{subsec:MAPPIES__SeedPhotons}
The seed photons are drawn in the laboratory frame from either an isotropic, single-temperature black body distribution, or from a multi-temperature accretion-disk spectrum. In the first case, the polar angle $\Theta_{i,lab}$ and azimuthal angle $\Phi_{i,lab}$ is drawn from an isotropic distribution, so that
\begin{eqnarray}
\Theta_{i,lab} &=& \mathrm{cos}^{-1}(2\xi_1 -1) \nonumber \\
\Phi_{i,lab} &=& 2\pi\xi_2
\end{eqnarray}
with $\xi_1$ and $\xi_2$ are random numbers between 0 and 1. The seed photon energy $\epsilon_{i,lab}$ is drawn from a black body distribution that corresponds to a single temperature $kT_{rad}$ (defined as an input parameter), following the Monte-Carlo methods by \cite{Pozdnyakov_etal1983}. The photon is then boosted to the co-moving frame of the emission region with the bulk boost equations \citep{Bottcher_etal2012}
\begin{eqnarray}
\label{eq:boostEM}
\epsilon_{i,em} &=& \Gamma_{jet} \epsilon_{i,lab} \left(1 - \beta_{jet}\cos \Theta_{i,lab}\right) \nonumber \\
\cos \Theta_{i,em} &=& \frac{\cos \Theta_{i,lab} - \beta_{jet}}{1 - \beta_{jet} \cos \Theta_{i,lab}}.
\end{eqnarray}

\begin{figure}[ht!]
\plotone{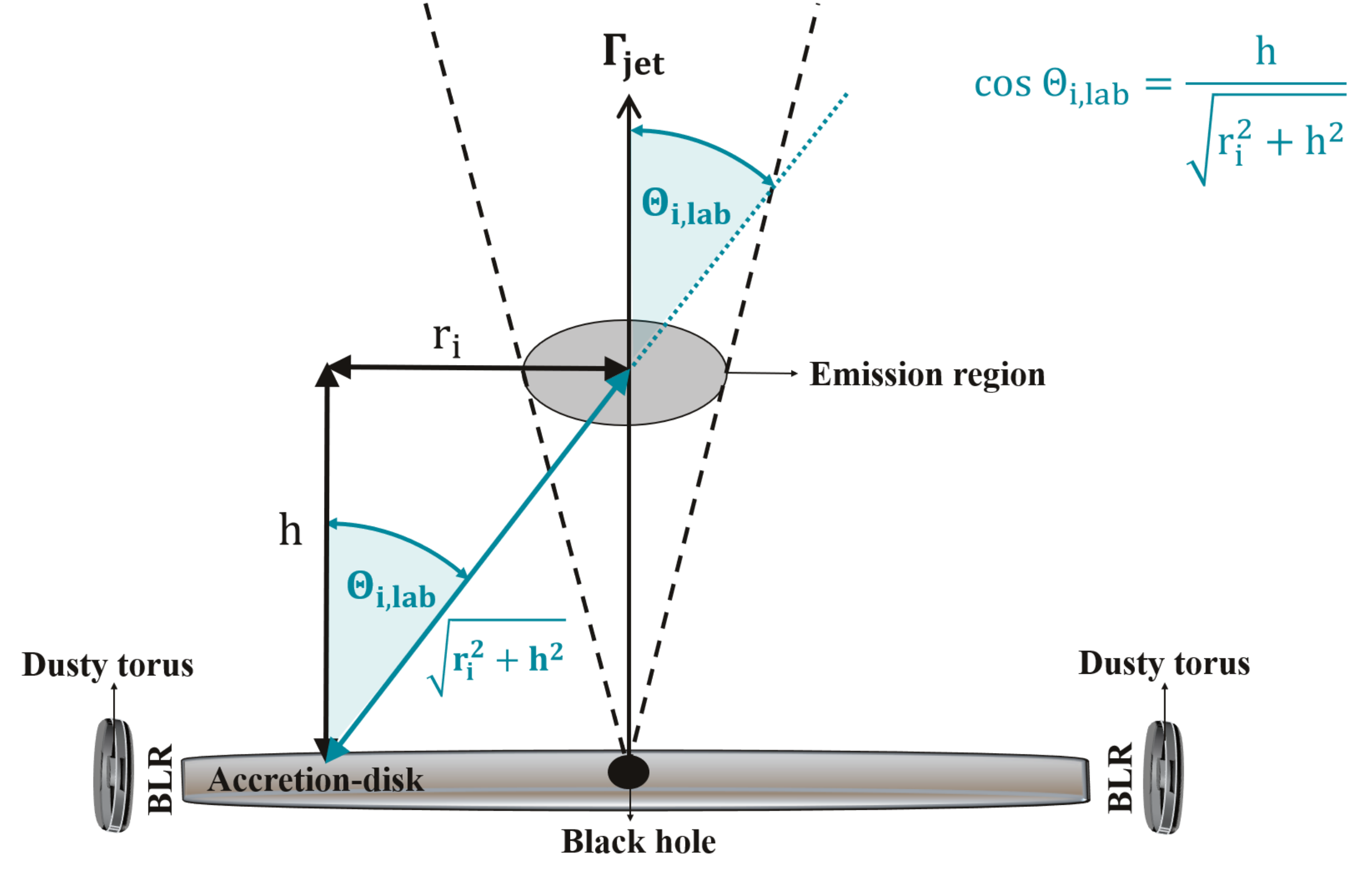}
\caption{An illustration of how a single photon (denoted with a subscript \textit{i}) is drawn from an accretion-disk spectrum. The angle between the line of sight from the emission region $r_i$ and the normal to the disk is denoted with $\Theta_{i,lab}$, and $h$ is the distance between the central black hole and the emission region. \label{fig:ADGeo}}
\end{figure}

An illustration of how a single photon is drawn from an accretion-disk spectrum is given in Figure \ref{fig:ADGeo}. The calculation requires the following input parameters: The mass of the black hole $M_{BH}$, the inner-disk radius $R_D^{in}$, the outer-disk radius $R_D^{out}$, the accretion-disk luminosity $L_D$, and the height of the emission region $h$ (i.e. the distance between the central black hole and the emission region). The flux per unit radius is given by
\begin{eqnarray}
\label{eq:flux}
dF = \left(2\pi r \cos \Theta_{lab} \sigma_{SB} T_{rad}^4(r)\right) dr
\end{eqnarray}
where
\begin{eqnarray}
\label{eq:radial_structure}
kT_{rad}(r) = k\left[\frac{3GM_{BH}\dot{m}}{8\pi r^{3} \sigma_{SB}}\left(1 - \sqrt{\frac{R_D^{in}}{r}}\right)\right]^{\frac{1}{4}}
\end{eqnarray}
is the radial temperature structure, and
\begin{eqnarray}
\label{eq:mu}
\cos \Theta_{lab} = \frac{h}{\sqrt{r^2 + h^2}}
\end{eqnarray}
is the cosine of the angle between the line of sight from the emission region to $r$ and the normal to the disk. In Equation \ref{eq:radial_structure}, $\dot{m} = \dot{M}_{Edd} (L_D/L_{Edd})$ is the accretion rate of the disk, with $\dot{M}_{Edd} = L_{Edd}/(0.1c^2)$ the Eddington accretion rate relative to the Eddington luminosity $L_{Edd} = [(1.26 \times 10^{46}M_{BH})/(10^8 M_{\odot})]~\mathrm{erg \cdot s^{-1}}$, $\sigma_{SB} \sim 5.7 \times 10^{-5}\mathrm{erg \cdot cm^{-2}\cdot s^{-1} \cdot K^{-4}}$ is the Stefan-Boltzmann constant, $G \sim 2\times 10^{-8}\mathrm{cm^3 \cdot g^{-1} \cdot s^{-2}}$ is the gravitational constant, $c \sim 3 \times 10^{10}\mathrm{cm \cdot s^{-1}}$ is the speed of light, and $M_{\odot} \sim 2\times 10^{33}$ g is the solar mass.

The cumulative probability to receive a photon from a radius $r < r_i$ is obtained from Equation \ref{eq:flux} as 
\begin{eqnarray}
\label{eq:Pr}
P(r < r_i) &=& \frac{1}{N} \int^{r_i}_{R_D^{in}} \left[\frac{1}{r^2\sqrt{r^2 + h^2}} - \frac{\sqrt{R_D^{in}}}{r^{\frac{5}{2}}\sqrt{r^2 + h^2}}\right] dr \nonumber \\
&=& N^{-1} (I_1 - \sqrt{R_D^{in}} I_2)
\end{eqnarray}
with the normalisation given by 
\begin{eqnarray}
N = \int^{R_D^{out}}_{R_D^{in}} \left[\frac{1}{r^2\sqrt{r^2 + h^2}} - \frac{\sqrt{R_D^{in}}}{r^{\frac{5}{2}}\sqrt{r^2 + h^2}}\right] dr.
\end{eqnarray}
In Equation \ref{eq:Pr},  
\begin{eqnarray}
I_1 &=& \int^{r_i}_{R_D^{in}} \frac{1}{r^2\sqrt{r^2 + h^2}} dr \nonumber \\
&=& \left[ - \frac{\sqrt{r^2 + h^2}}{h^2r}\right]^{r_i}_{R_D^{in}} 
\end{eqnarray}
and 
\begin{eqnarray}
\label{eq:I2}
I_2 = \int^{r_i}_{R_D^{in}} \frac{1}{r^{\frac{5}{2}}\sqrt{r^2 + h^2}} dr
\end{eqnarray}
which is solved with the assumption that  $r \ll h$ in most relevant cases. The typical outer-disk radius $R_D^{out} \sim 10^3 R_G 10^3 \approx 1.5 \times 10^{16} M_8$ cm, where $M_8 = M_{BH}/(10^8 M_{\odot})$ and $R_G$ is the gravitational radius of the black hole. Therefore, $r < h$ anywhere in the disk when $M_{BH} \lesssim 3 \times 10^{8} M_\odot$ and/or $h \lesssim 10^{17}$. The Taylor expansion,
\begin{eqnarray}
\frac{1}{\sqrt{r^2 + h^2}} &=& \frac{1}{z \sqrt{1 + (r/z)^2}} \nonumber\\
&=& \frac{1}{z} \sum^\infty_{n=0} \binom{-\frac{1}{2}}{n} \left(\frac{r}{z}\right)^{2n}
\end{eqnarray}
can then be used, so that, to first order
\begin{eqnarray}
I_2 = \sum_{n=0}^{\infty} \binom{-\frac{1}{2}}{n} z^{-(2n+1)} \int^{r}_{R_D^{in}} r^{2n-\frac{3}{2}} dr
\end{eqnarray}
For typical emission-region heights of $h \gtrsim 10^{17}$~cm, $r \ll h$ everywhere in the disk, which gives
\begin{eqnarray}
I_2(r \ll h) = \left[ - \frac{2}{hr^{\frac{3}{2}}} - \frac{\sqrt{r}}{h^3}\right]^{r_i}_{R_D^{in}}
\end{eqnarray}
The radius $r_i$ is drawn by calculating $P(r < r_i)$ for different values of $r_i \in [R_D^{in}, R_D^{out}]$, until a given random number $\xi = P(r < r_i)$. The temperature of the disk at $r_i$, from which the photon energy is sampled, $KT_{i,rad}$, and polar angle of the seed photon, $\Theta_{i,lab}$, are subsequently calculated with Equations \ref{eq:radial_structure} and \ref{eq:mu}, respectively. The energy of the photon $\epsilon_{i,lab}$ is drawn from a black body distribution that corresponds to $KT_{i,rad}$ (following the Monte-Carlo methods by \cite{Pozdnyakov_etal1983}), and boosted to the co-moving frame of the emission region with Equation \ref{eq:boostEM}.
\subsection{Electron energy distributions}\label{subsec:MAPPIES__Electron}
The electrons are assumed to be isotropic in the co-moving frame of the emission region. An electron is randomly assigned to every photon as it is transported through the computational domain, drawn from (1) a purely thermal (Maxwell) distribution, (2) a hybrid (Maxwell + power-law) distribution, or (3) a \textit{user-defined} input electron spectrum. In the first case, the thermal temperature $kT_{e}$ is given as an input parameter, and a Lorentz factor $\gamma_{i}$ of an electron is drawn from a Maxwellian distribution, following the Monte-Carlo methods by \cite{Pozdnyakov_etal1983}. In the case of thermal and a power-law tail of non-thermal electrons three additional parameters are required: The fraction of the electrons that are assumed to be non-thermal $f_{nth}$, the power-law index of the non-thermal electrons $p$ (which are drawn from a power-law distribution $n_{pl}(\gamma) = N_{pl} \gamma^{-p}$, with $N_{pl}$ the normalisation constant), and the Lorentz factor that corresponds to the cut-off energy of the power-law tail $\gamma_{max}$. The Lorentz factor which corresponds to where the power-law tail begins, $\gamma_{min}$, is determined by iteration until 
\begin{eqnarray} \label{eq:fnth} 
f_{nth} = \frac{n_{pl}}{n_{th} + n_{pl}} 
\end{eqnarray}
is equal to the input parameter $f_{nth}$ defined, where 
\begin{eqnarray}
n_{pl} = N_{pl} \int^{\gamma_{max}}_{\gamma_{min}} \tilde{\gamma}^{-p} d\tilde{\gamma}
\end{eqnarray}
is the power-law tail and
\begin{eqnarray}
n_{th} = N_{th} \int^{\gamma_{min}}_{1} \tilde{\gamma}^{2} \tilde{\beta} e^{-\frac{\tilde{\gamma}}{\Theta_e}} d\tilde{\gamma}
\end{eqnarray}
is the number of thermal electrons, with $\tilde{\beta} = \sqrt{1 - \tilde{\gamma}^{-2}}$, $\Theta_e = kT_{e}/(m_ec^2)$, and $N_{th} = 1$ an arbitrary normalisation constant. At the point of intersection,
\begin{eqnarray}
N_{pl} \gamma_{min}^{-p} = N_{th} \gamma_{min}^2\beta_{min} e^{-\frac{\gamma_{min}}{\Theta_e}}
\end{eqnarray}
During the Monte-Carlo simulation, a random number $\xi_1 \in [0,1]$ is compared to $f_{nth}$ in order to draw $\gamma_{i}$ from either a power-law distribution or a Maxwellian distribution. If $\xi_1 \leq f_{nth}$, $\gamma_i$ is drawn from a power-law distribution,
\begin{eqnarray}
\gamma_i = \left[\xi\left( \gamma_{max}^{1-p} - \gamma_{min}^{1-p} \right) + \gamma_{min}^{1-p}\right]^{\frac{1}{1-p}}
\end{eqnarray}
where $\xi \in [0,1]$. If $\xi_1 > f_{nth}$, $\gamma_i$ is drawn from a Maxwellian distribution, in which case, a new random number $\xi_1 \in [0,1]$ will be drawn if $\gamma_i > \gamma_{min}$, and the process will be repeated.

In the case where the electron energy is drawn from a \textit{user-defined} distribution, an input electron spectrum $n(\gamma)$ alone is required. The cumulative distribution function (CDF),
\begin{eqnarray}
\label{inputElectrons}
P(\gamma < \gamma_i) = \frac{\int^{\gamma_i}_1 n(\tilde{\gamma}) d\tilde{\gamma}}{\int^{\gamma_{max}}_1 n(\tilde{\gamma}) d\tilde{\gamma}}
\end{eqnarray}
is calculated from the input electron spectrum, and $\gamma_{i}$ is drawn such that $\xi = P(\gamma < \gamma_{i})$.
\subsection{Polarisation-dependent Compton Scattering}\label{subsec:MAPPIES__Compton}
The polarisation signatures of the photons are calculated using the Stokes formalism \citep{Stokes_1851}. The contributions of the photon to the second ($Q_i$) and third ($U_i$) Stokes parameters are calculated by following the Monte-Carlo methods by \cite{Matt_etal1996}, and summed over all the photons after the simulation is complete. Since the photons are individually transported through the computational domain, every photon is $100\%$ polarised. However, the external radiation is assumed to be unpolarised. The polarisation vector $\vec{P}_{i,em}$ (which points in the direction of the electric field vector) of every seed photon is thus randomly drawn perpendicular to the photon's direction of propagation, which results in a zero net polarisation (see e.g. \cite{Tamborra_etal2018}). 

The probability of the photon to undergo Compton scattering is determined by the (polarisation averaged) Compton cross section, 
\begin{eqnarray}
\label{eq:KN}
\frac{\sigma_{KN}}{\sigma_T} &=& \frac{3}{4} \left[ \frac{1 + x_e}{x_e^3} \left( \frac{2x_e(1 + x_e)}{1 + 2x_e} - \ln (1 + 2x_e)\right) \right] \nonumber \\
&~& + \frac{3}{4} \left[ \frac{1}{2x_e} \ln(1 + 2x_e) - \frac{1 + 3x_e}{\left(1 + 2x_e\right)^2}\right]
\end{eqnarray}
most conveniently evaluated in the electron rest frame, with $\sigma_T \sim 6.7 \times 10^{-25} ~\mathrm{cm^2}$ the Thomson cross section \citep{Bottcher_etal2012}, and  $x_{e} = \epsilon_{e}/m_ec^2$ the dimensionless seed photon energy. The seed photon is therefore transformed to the electron rest frame with the Lorentz matrix \citep{Bonometto_etal1970, Tamborra_etal2018}, and a random number $\xi \in [0,1]$ is drawn to determine whether the seed photon will be scattered.  If $\xi > (\sigma_{i,KN}/\sigma_T)$ (where $\sigma_{i,KN}$ is the full Klein-Nishina cross section for the current photon), the photon will continue in the same direction without scattering, otherwise the Compton scattering event will be simulated. 

\begin{figure}[ht!]
\plotone{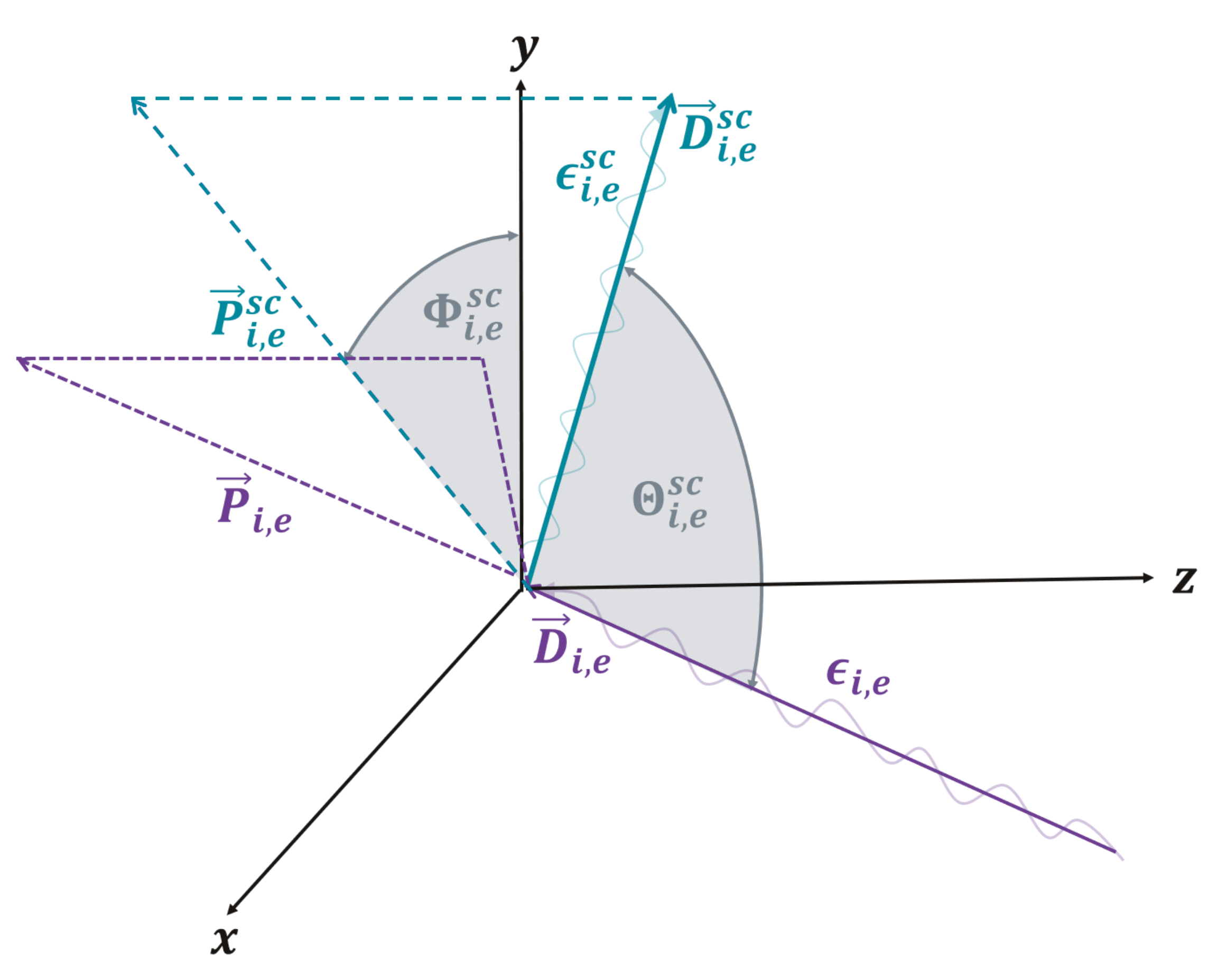}
\caption{Illustration of the geometry of the Compton effect for a single photon (denoted with a subscript \textit{i}) in the electron rest frame. The seed photon, moving in a direction $\vec{D}_{i,e}$, with an energy $\epsilon_{i,e}$, and a polarisation vector $\vec{P}_{i,e}$ (which points in the direction of the electric field vector) is shown in purple. The scattered photon is shown in blue with a direction $\vec{D}^{sc}_{i,e}$, an energy $\epsilon^{sc}_{i,e}$, and a polarisation vector $\vec{P}^{sc}_{i,e}$. The polar scattering angle $\Theta^{sc}_{i,e}$ is the angle between the seed and scattered photon, while the azimuth scattering angle $\Phi^{sc}_{i,e}$ is the angle between $\vec{P}_{i,e}$ and the plane of scattering. \label{fig:ComptonGeo}}
\end{figure}

The geometry of the Compton effect for a single photon in the electron rest frame is illustrated in Figure \ref{fig:ComptonGeo}. The energy of the scattered photon is given by
\begin{eqnarray}
\label{eq:ICeps}
\epsilon_{i,e}^{sc} = \frac{\epsilon_{i,e}}{ 1 - x_{i,e} \cos \Theta_{i,e}^{sc}}
\end{eqnarray}
The probability of a photon to have a scattering angle of $\Theta_{i,e}^{sc}$ (the angle between the seed and scattered photon) is given by
\begin{eqnarray}
\label{eq:probTheta}
P(\Theta_{i,e}^{sc} < \Theta_{e}^{sc}) &=& \frac{x_{i,e}\left[ \frac{3}{2} + \mu_{e}^{sc}(1 - \frac{1}{2}\mu_{e}^{sc})\right] + \frac{1}{3}\left[1 + (\mu_{e}^{sc})^3 \right]}{\frac{2}{3} + 2x_{i,e} + \frac{1}{x_{i,e}} \ln (1 + 2x_{i,e})}
\nonumber \\
&-&\frac{\frac{1}{x_{i,e}}\bigg\{\ln \left[ 1 + x_{i,e}(1 - \mu_{e}^{sc})\right] - \ln( 1 + 2 x_{i,e})\bigg\}}{\frac{2}{3} + 2x_{i,e} + \frac{1}{x_{i,e}} \ln (1 + 2x_{i,e})} \nonumber \\
\end{eqnarray}
where $\mu_{e}^{sc} = \cos \Theta_{e}^{sc}$ \citep{Matt_etal1996}. The scattering angle $\Theta_{i,e}^{sc}$ is thus calculated by drawing a random number $\xi$ and finding the value of $\Theta_{e}^{sc}$ for which $P(\Theta_{i,e}^{sc} < \Theta_{e}^{sc}) = \xi$.

The azimuthal distribution of the photons is dependent on the seed photon polarisation. When the seed photons are unpolarised, the angle between $\vec{P}_{i,e}^{sc}$ and the plane of scattering, $\Phi_{i,e}^{sc}$, can be assumed to be isotropically distributed in the electron rest frame \citep{Matt_etal1996}. However, while following a \textit{single} photon approach, every individual photon is fully polarised. The azimuthal angle $\Phi_{i,e}^{sc}$ is thus drawn by calculating the probability of a fully polarised photon to have an angle $\Phi_{i,e}^{sc}$ \citep{Matt_etal1996},
\begin{eqnarray}
\label{eq:probPhi}
P(\Phi_{i,e}^{sc} < \Phi_{e}^{sc}) &=& \frac{1}{2\pi} \left[\Phi_{e}^{sc} - \frac{\sin^2 \Theta_{e}^{sc} \sin \Phi_{e}^{sc} cos \Phi_{e}^{sc}}{ \frac{x_{i,e}}{x_{i,e}^{sc}} + \frac{x_{i,e}^{sc}}{x_{i,e}^{sc}} - \sin^2\Theta_{e}^{sc}}\right]\nonumber\\
\end{eqnarray}
and finding the value of $\Phi^{sc}_e$ for which $P(\Phi^{sc}_{i,e} < \Phi^{sc}_e) = \xi$, where $\xi$ is a newly drawn random number.

The PD due to scattering is then calculated for a fully polarised photon as
\begin{eqnarray}
\label{eq:PD_due_scatt}
\mathrm{PD}_{i,e}^{sc} = 2\left[ \frac{1 - \sin^2\Theta^{sc}_{i,e}\cos^2\Phi^{sc}_{i,e}}{\frac{x_{i,e}}{x_{i,e}^{sc}} + \frac{x_{i,e}^{sc}}{x_{i,e}} - 2\sin^2\Theta^{sc}_{i,e}\cos^2\Phi^{sc}_{i,e}}\right]
\end{eqnarray}
 in the electron rest frame \citep{Matt_etal1996}, which determines whether the photon will be polarised after scattering. A random number $\xi \in [0,1]$ is drawn and compared to $\mathrm{PD}_{i,e}^{sc}$ in order to calculate the polarisation vector $\vec{P}_{i,e}^{sc}$ of the scattered photon.  If $\xi < \mathrm{PD}_{i,e}^{sc}$, the photon will contribute to the partially polarised Compton emission with a polarisation vector of
 \begin{eqnarray}
 \vec{P}^{sc}_{i,e} = \frac{(\vec{P}_{i,e} \times \vec{D}^{sc}_{i,e}) \times \vec{D}^{sc}_{i,e}}{|\vec{P_{i,e}^{sc}}|}
 \end{eqnarray}
where $\vec{D}^{sc}_{i,e}$ is the scattered photon's direction of propagation \citep{Angel_1969}. Otherwise, $\vec{P}_{i,e}^{sc}$ is randomly drawn perpendicular to $\vec{D}_{i,e}^{sc}$, and will contribute to the unpolarised fraction of the Compton emission \citep{Matt_etal1996, Tamborra_etal2018}.

The scattered photon is transformed back into the co-moving frame of the emission region with the Lorentz matrix, where the photon's contribution to the second ($Q^{sc}_i$) and third ($U^{sc}_i$) Stokes parameters is calculated. The photon is boosted into the laboratory frame with the bulk boost equations \citep{Bottcher_etal2012}, 
\begin{eqnarray}
\epsilon_{i,lab}^{sc} &=& \Gamma_{jet} \epsilon_{i,em}^{sc} \left(1 + \beta_{jet}\cos \Theta_{i,em}^{sc}\right) \nonumber \\
\cos \Theta_{i,lab}^{sc} &=& \frac{\cos \Theta_{i,em}^{sc} + \beta_{jet}}{1 + \beta_{jet} \cos \Theta_{i,em}^{sc}}
\end{eqnarray}
and shifted to the observer's frame where $\epsilon_{i,obs}^{sc} = \epsilon_{i,lab}^{sc}/(1 + z)$, with $z$ the redshift of the source.

The Compton polarisation signatures are calculated after the simulation is complete by summing the contributions of the scattered photons to the Stokes parameters, so that
\begin{eqnarray}
\mathrm{Q} = \sum_{i=0}^{N^{sc}_{phot}} \mathrm{Q^{sc}_i} \quad and \quad \mathrm{U}= \sum_{i=0}^{N^{sc}_{phot}} \mathrm{U^{sc}_i}
\end{eqnarray}
where $N^{sc}_{phot}$ is the number of the scattered photons in the specified direction. The polarisation signatures of the Compton emission are determined as
\begin{eqnarray}
\mathrm{PD} = \frac{\sqrt{\mathrm{Q}^2 + \mathrm{U}^2}}{N^{sc}_{phot}} \quad and \quad \mathrm{PA} = \frac{1}{2} \tan^{-1} \frac{\mathrm{U}}{\mathrm{Q}}
\end{eqnarray}
and binned in viewing angles, $\Theta_{i,lab}^{sc}$, and energy, $\epsilon_{i,obs}^{sc}$. This allows us to identify the viewing angle and photon energy range at which the maximum PD occurs, thus offering the best opportunities to measure Compton polarisation.
\section{Compton polarisation in the Thomson and Klein-Nishina regimes}\label{sec:MSC_Results}
\begin{figure*}[ht!]
\plotone{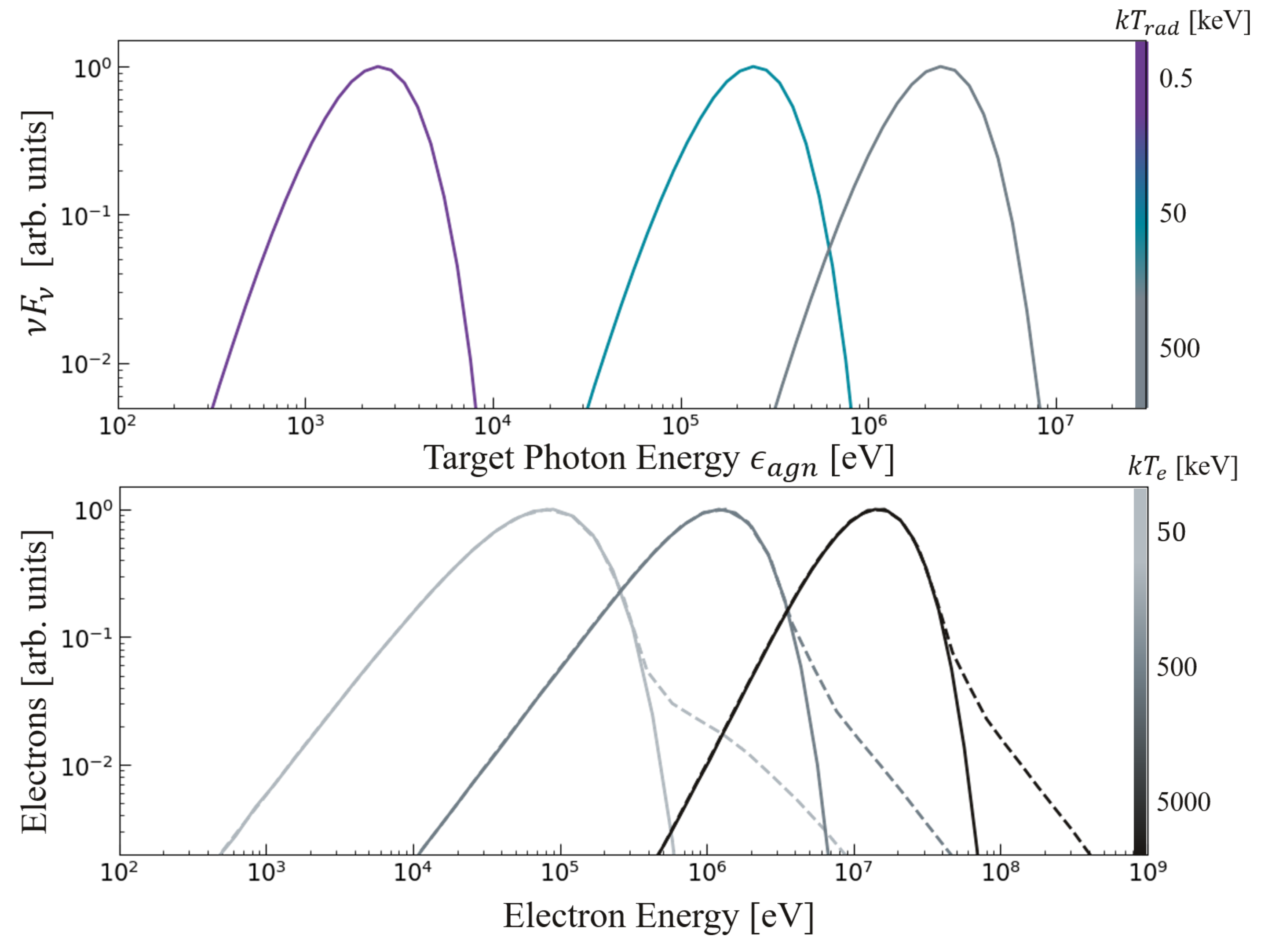}
\caption{The seed photon spectra (top panel) and electron energy distributions (bottom panel). The photons (top panel) are drawn in the laboratory frame from an isotropic, single-temperature black body distributions with $kT_{rad} = 0.5$ keV (shown in purple), $kT_{rad} = 50$ keV (shown in blue), and $kT_{rad} = 500$ keV (shown in grey). The electrons (bottom panel) are isotropic in the co-moving frame of the emission region with thermal temperatures of $kT_e = 50$ keV, $kT_e = 500$ keV, and $kT_e = 5000$ keV that increase with the shade of grey in the bottom panel. The electron energy is drawn from either a purely thermal distribution (shown with solid lines) or a hybrid (Maxwell + power-law) distribution (shown with dashed lines) for the combination of free parameters listed in Table \ref{tab:app_param}.  \label{fig:app_input}}
\end{figure*}

The MAPPIES code presented in this paper can be used to simulate the polarisation due to Compton scattering of different seed photon fields and electrons with different energy distributions. In this section, generic results for the Compton polarisation in the Thomson and Klein-Nishina regimes are presented. Only results for the isotropic black body target photons are shown in this paper, while results for an accretion-disk spectrum will be presented in a companion paper (Dreyer and B{\"o}ttcher 2020, in preparation) for applications to specific AGN. The results are shown for the combination of free parameters listed in Table \ref{tab:app_param}. The seed photons are drawn in the laboratory frame from an isotropic, single-temperature black body distribution (shown in the top panel of Figure \ref{fig:app_input}) with $kT_{rad} = 0.5$ keV, $kT_{rad} = 50$ keV, and $kT_{rad} = 500$ keV. The electrons (shown in the bottom panel of Figure \ref{fig:app_input}) are assumed to be isotropic in the co-moving frame of the emission region (that moves along the jet with a bulk Lorentz factor of $\Gamma_{jet} = 10$) with thermal temperatures of $kT_e = 50$ keV, $kT_e = 500$ keV, and $kT_e = 5000$ keV. The electron energy is drawn from either a purely thermal distribution (shown with solid lines) or a hybrid (Maxwell + power-law) distribution (shown with dashed lines).  In all the figures discussed, results for soft $X$-rays ($kT_{rad} = 0.5$ keV) are shown in purple, results for hard $X$-rays ($kT_{rad} = 50$ keV) are shown in blue, and results for $\gamma$-rays ($kT_{rad} = 500$ keV) are shown in grey.
\begin{deluxetable}{lc}
\tablenum{3}
\tablecaption{The input parameters for the generic results of Compton polarisation in the Thomson and Klein-Nishina regimes. \label{tab:app_param}}
\tablewidth{0pt}
\tablehead{\colhead{\textbf{Input Parameter}} & \colhead{Value}}
\startdata
\hline
Lorentz factor of the jet, $\Gamma_{jet}$ & $10$ \\
\hline
Number of seed photons considered & $10^8$ \\
\hline
Temperature of the seed photons, $kT_{rad}$ & $0.5$ keV\\ & $50$ keV\\
& $500$ keV\\
\hline
Thermal temperature of electrons, $kT_{e}$ &  $50$ keV\\
& $500$ keV\\
& $5000$ keV\\
\hline
Fraction of non-thermal electrons, $f_{nth}$ & $0.02$ \\
\hline
Power-law index of the power-law & \\
distribution of non-thermal electrons, $p$ & $2.0$\\
\hline
Maximum Lorentz factor of the power-law & \\
distribution of non-thermal electrons, $\gamma_{max}$ & $1.6\times10^3$ \\
\enddata
\end{deluxetable}

\subsection{The Compton spectra}\label{subsec:ComptonSpectra}
\begin{figure*}[ht!]
\plotone{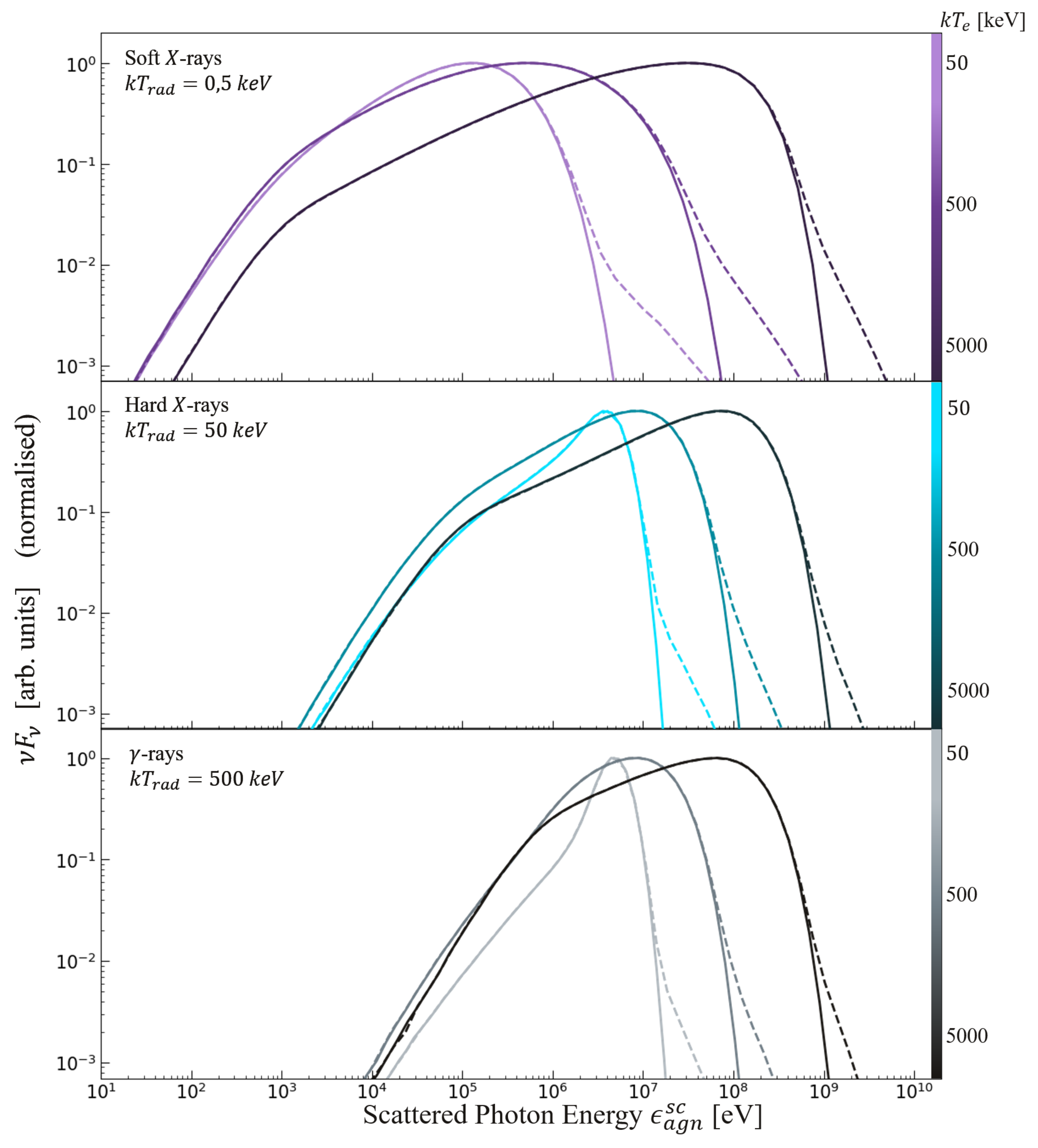}
\caption{The external Compton spectrum for soft $X$-rays ($kT_{rad} = 0.5$ keV; shown in purple), hard $X$-rays ($kT_{rad} = 50$ keV; shown in blue), and $\gamma$-rays ($kT_{rad} = 500$ keV; shown in grey). The results are shown for scattering off non-relativistic ($kT_e = 50$ keV), mildly-relativistic ($kT_e = 500$ keV), and relativistic ($kT_e = 5000$ keV) electrons with thermal temperatures that increases with the shade of color in each panel. The electrons are drawn from either a purely thermal distribution (shown with solid lines) or a hybrid (Maxwell + power-law) distribution (shown with dashed lines) for the combination of free parameters listed in Table \ref{tab:app_param}.  \label{fig:app_Spectrum}}
\end{figure*}

The external Compton spectra resulting from the different combinations of target radiation fields and electron distributions mentioned above, are shown in Figure \ref{fig:app_Spectrum}. Compton scattering in the Thomson regime ($x_e \ll 1$) is almost elastic in the electron rest frame, while the energy exchange between the seed photon and electron becomes substantial, along with a reduction of the cross section, in the Klein-Nishina regime ($x_e \gg 1$). Due to relativistic boosting, the photons that are scattered in the Thomson regime have energies $\epsilon_{lab}^{sc} \sim \gamma^2 \Gamma_{jet}^2 \epsilon_{lab}$ (where $\gamma$ is the averaged Lorentz factor of the electrons) in the laboratory frame. Compton scattering off a power-law distribution of non-thermal electrons results in a power-law distribution of scattered photons (shown with dashed lines). The photons that are scattered in the Klein-Nishina regime have cut-off energies in the laboratory frame that correspond to the reduction of the cross section in the electron rest frame. For electrons with thermal energies of $kT_e = 50$ keV, all the electrons are non-relativistic. Soft $X$-rays and hard $X$-rays are thus scattered in the Thomson regime with energies $\sim \gamma^2 \Gamma_{jet}^2$ higher than the seed photon energies. For mildly-relativistic electrons, the peak of the electron distribution is at $\gamma \sim 2$. Hard $X$-rays and $\gamma$-rays are thus scattered in the Klein-Nishina regime with very similar high-energy spectra (shown in blue and grey in Figure \ref{fig:app_Spectrum}). 

\subsection{The polarisation degree in the Thomson and Klein-Nishina regimes}
The Compton cross section is generally dependent on polarisation. The most familiar form of the polarisation-dependent differential cross section is given by 
\begin{eqnarray}
\label{eq:diffKNpol}
\frac{d\sigma_{KN}}{d\Omega_{e}^{sc}} = \frac{1}{4}r_e^2 \left(\frac{x_e}{x_e^{sc}}\right)^2 \left[\frac{x_e}{x_e^{sc}} + \frac{x_e^{sc}}{x_e} - 2 + 4\cos^2\theta \right] 
\end{eqnarray}
in the electron rest frame, where $\theta$ is the angle between the polarisation vector of the seed photon $\vec{e}_e$ and the polarisation vector of the scattered photon $\vec{e}_e^{sc}$ \citep{Matt_etal1996}. From Equation \ref{eq:diffKNpol}, 
\begin{eqnarray}
\label{eq:sigamPOL}
\frac{d\sigma_{KN}}{d\Omega_{e}^{sc}} \propto \frac{x_e}{x_e^{sc}} + \frac{x_e^{sc}}{x_e} - 2 + 4(\vec{e}_e \cdot \vec{e}_e^{sc}).
\end{eqnarray}
 Since the polarisation term in Equation \ref{eq:sigamPOL} dominates for $x_e \ll 1$, photons that are scattered in the Thomson regime are expected to be polarised. The polarisation term becomes negligible for $x_e \gg 1$, and polarisation is thus not expected to be induced for Compton scattering in the Klein-Nishina regime.

\begin{figure*}[ht!]
\plotone{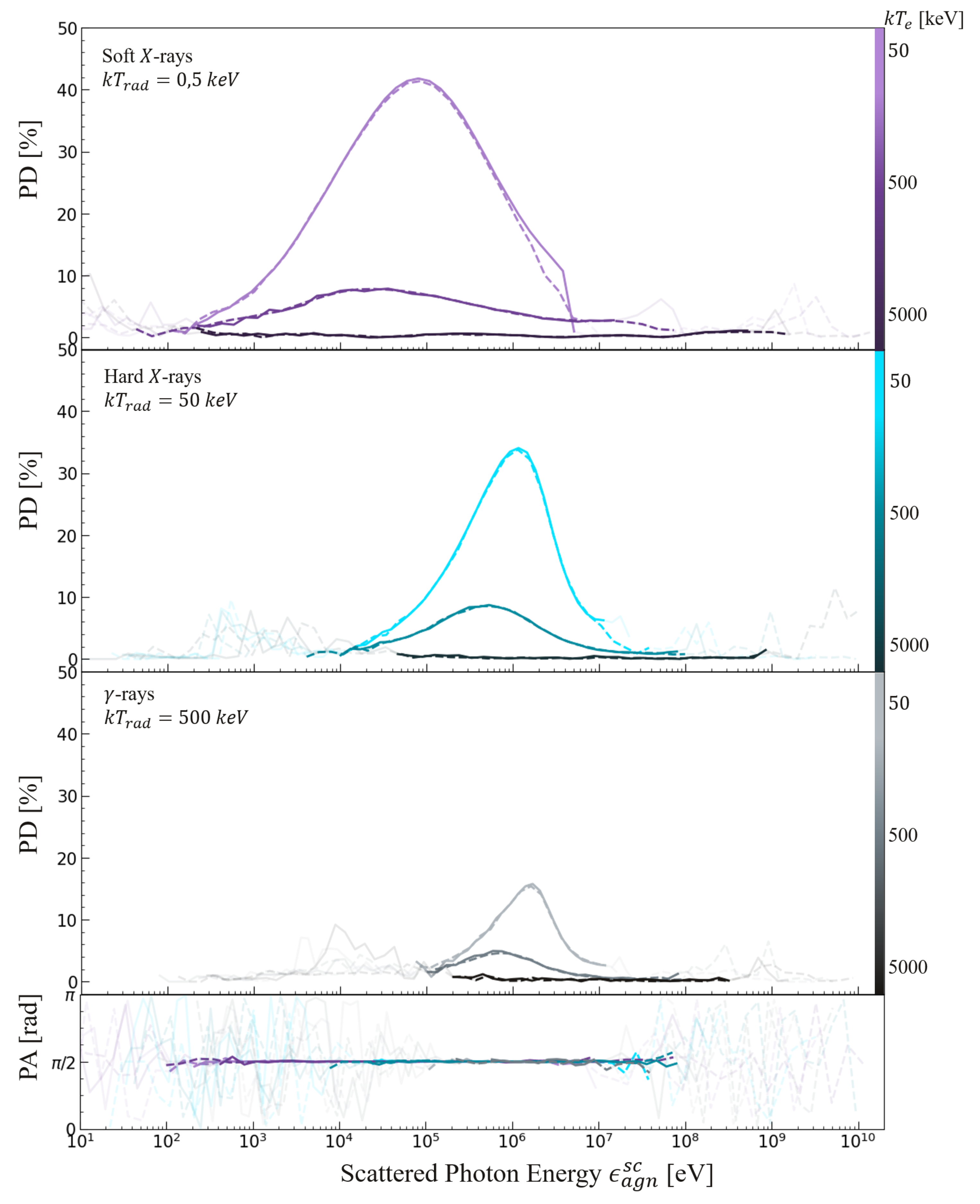}
\caption{The polarisation signatures as a function of the scattered photon energy for soft $X$-rays ($kT_{rad} = 0.5$ keV; shown in purple), hard $X$-rays ($kT_{rad} = 50$ keV; shown in blue), and $\gamma$-rays ($kT_{rad} = 500$ keV; shown in grey). The results are shown for scattering off non-relativistic ($kT_e = 50$ keV), mildly-relativistic ($kT_e = 500$ keV), and relativistic ($kT_e = 5000$ keV) electrons with thermal temperatures that increases with the shade of color in each panel. The electrons are drawn from either a purely thermal distribution (shown with solid lines) or a hybrid (Maxwell + power-law) distribution (shown with dashed lines) for the combination of free parameters listed in Table \ref{tab:app_param}.  \label{fig:PolvsEnergy}}
\end{figure*}

The polarisation signatures are shown as a function of the scattered photon energy $\epsilon_{lab}^{sc}$ in Figure \ref{fig:PolvsEnergy}, for viewing angles of $\Theta_{lab}^{sc} \sim \Gamma_{jet}^{-1}$ rad. The PD decreases with increasing photon energies and further decreases for higher electron temperatures, because the polarisation arises only for photons that are scattered in the Thomson regime. The maximum PD for Compton emission in the Thomson regime occurs where the thermal, non-relativistic electrons scatter the seed photons (i.e. to an energy of $\Gamma_{jet}^2 kT_{rad}$). For electron temperatures of $kT_e = 5\times10^3$ keV, all the electrons are highly relativistic with $\gamma \gtrsim 10$, and no Compton polarisation is induced, irrespective of whether the photons are scattered in the Thomson or Klein-Nishina regime. There are more photons produced at higher energies for scattering off a power-law distribution of non-thermal electrons (shown with dashed lines) than in the case of scattering off purely thermal electrons (shown with solid lines), but those photons are unpolarised since the non-thermal electrons are relativistic.

\subsection{The polarisation signatures in the Thomson regime}
\begin{figure*}[ht!]
\plotone{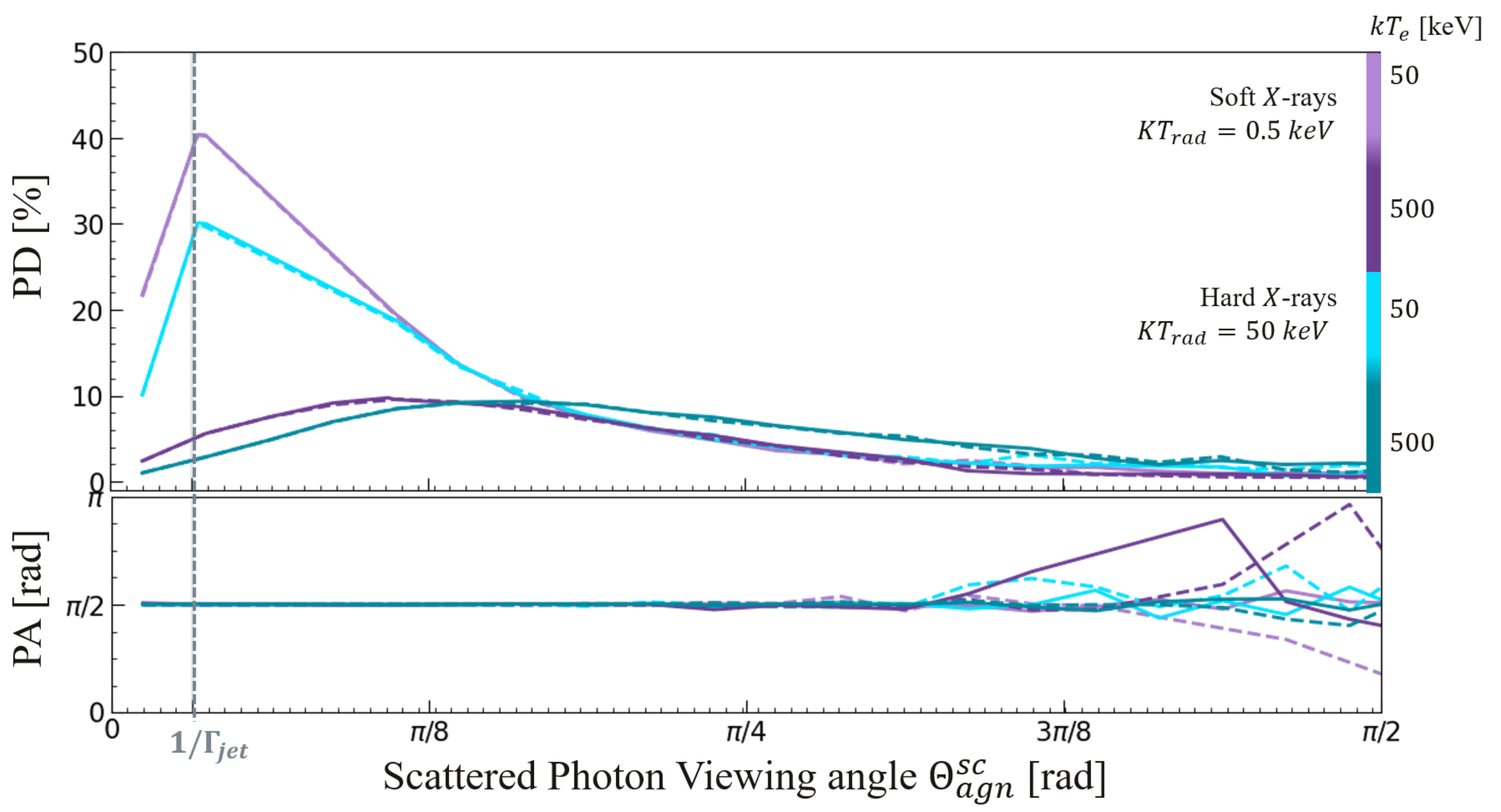}
\caption{The polarisation signatures as a function of the scattered photon viewing angle for soft $X$-rays ($kT_{rad} = 0.5$ keV; shown in purple) and hard $X$-rays ($kT_{rad} = 50$ keV; shown in blue). The results are shown for scattering off non-relativistic ($kT_e = 50$ keV) and mildly-relativistic ($kT_e = 500$ keV) electrons with thermal temperatures that increases with the shade of purple and blue. The electrons are drawn from either a purely thermal distribution (shown with solid lines) or a hybrid (Maxwell + power-law) distribution (shown with dashed lines) for the combination of free parameters listed in Table \ref{tab:app_param}. The grey line indicates the viewing angle of $\Theta_{lab}^{sc} = \Gamma_{jet}^{-1}$~rad, where $\Gamma_{jet} = 10$.  \label{fig:PolvsAngle}}
\end{figure*}

The polarisation signatures due to scattering off non-relativistic and mildly-relativistic electrons are given as a function of the scattered photon viewing angle $\Theta_{lab}^{sc}$ in Figure \ref{fig:PolvsAngle} (averaged over all photon energies). Since the orientation of the polarisation does not change significantly for different electron energy distributions, the polarisation signatures are similar for scattering off thermal electrons (shown with solid lines) and a power-law distribution of non-thermal electrons (shown with dashed lines). The PA is shown as a function of $\epsilon_{lab}^{sc}$ and $\Theta_{lab}^{sc}$ in the bottom panels of Figures \ref{fig:PolvsEnergy} and \ref{fig:PolvsAngle}, respectively. In both cases, the PA for Compton emission that are polarised assumes a constant value of PA $= \pi/2$~rad, that corresponds to polarisation perpendicular with respect to the jet.

The PD as a function of $\Theta_{lab}^{sc}$ (Figure \ref{fig:PolvsAngle}) shows at which angles the maximum polarisation occurs. For scattering to happen in the Thomson regime, the electrons have to move in the same direction as the seed photons (backwards in the jet). The scattered photons move perpendicular to their incoming direction in the electron rest frame which appears at an angle of $\sim \gamma^{-1}$~rad with respect to the backward direction. For photons that are scattered in the Thomson regime, the maximum PD occurs at a right angle of $\Theta_{e}^{sc} \sim \pi/2$~rad in the electron rest frame, which is essentially the same in the emission-region rest frame for scattering off non-relativistic ($kT_e = 50$~keV; when there are hardly any relativistic motions) electrons. Due to relativistic boosting, photons at $\Theta_{em}^{sc} \sim \pi/2$~rad in the emission-region rest frame will be observed at $\Theta_{lab}^{sc} \sim \Gamma_{jet}^{-1}$~rad in the laboratory frame. The maximum PDs for scattering off non-relativistic electrons occur thus at angles of $\Theta_{lab}^{sc} \sim \Gamma_{jet}^{-1}$~rad, indicated with a grey dashed line in Figure \ref{fig:PolvsAngle}. 

The peak of the electron distribution for mildly-relativistic ($kT_e = 500$~keV) electrons is around $\gamma \sim 2$. Soft $X$-rays with $kT_{rad} = 0.5$~keV are boosted to $\sim (0.5~\mathrm{keV})\Gamma_{jet} = 5.0$~keV into the emission frame, with the black body spectrum peaking at $x_{em} = \Gamma_{jet}\frac{2.8 kT_{rad}}{m_ec^2} \sim 0.03$. Similarly, hard $X$-rays with $kT_{rad} = 50$~keV are boosted to $\sim 500$~keV into the emission frame, with the black body spectrum peaking at $x_{em}\sim 2.7$. Therefore, photons that are scattered by mildly-relativistic electrons have angles of $\Theta_{em}^{sc} = (\pi - \gamma^{-1})$~rad in the emission-region rest frame. Relativistic aberration into the laboratory frame causes the maximum PD to occur at angles that are larger than those in the case of scattering off non-relativistic electrons. 
\section{Summary and conclusion}\label{sec:summary}
The MAPPIES code presented in this paper is capable of predicting the Compton polarisation in different jet-like astrophysical sources for different photon energies and electron temperatures (see Appendix \ref{sec:MSC_Compare} for a comparison to previously published results). The effects of Compton scattering depend on the temperature of the seed photons, as well as the Lorentz factors and energy distribution of the electrons. The photons scatter to higher energies (with a factor of $\Gamma_{jet}^2\gamma^2$) in the Thomson regime, and have cut-off energies that correspond to the reduction of the cross section in the Klein-Nishina regime. The PD of the scattered photons depends on the effects of Compton scattering due to the polarisation dependence of the Klein-Nishina cross section, given by Equation \ref{eq:diffKNpol}. The PD decreases with the increase of photon energies and higher electron temperatures. The energy regimes with non-negligible PDs shift to higher energies and become smaller for higher seed photon temperatures, while narrowing further for higher electron temperatures. Polarisation is therefore expected to be induced for photons that are scattered in the Thomson regime, and no polarisation is expected to be induced for photons that are scattered in the Klein-Nishina regime.  For electron temperatures of $kT_e = 5000$~keV,
essentially all the electrons are highly relativistic and no Compton polarisation is induced irrespective of whether the photons are scattered in the Thomson or Klein Nishina regime.  

The maximum PD for scattering in the Thomson regime occurs at viewing angles of $\Theta_{lab}^{sc} \sim \Gamma_{jet}^{-1}$~rad (shown in Figure \ref{fig:PolvsAngle}). The maximum PD occurs at larger angles for scattering off mildly-relativistic electrons, which suggests that Compton polarisation is sensitive to relativistic aberration for mildly-relativistic electrons with Lorentz factors of $\gamma \gtrsim 2$. The PA for the fraction of the scattered photons that are polarised assumes a constant value of PA $\sim \pi/2$~rad, which corresponds to polarisation perpendicular with respect to the jet, regardless of the photon energy and electron temperature.

In the view of the future proposed high-energy polarimetry missions listed in Table \ref{tab:PolMissions}, the MAPPIES code can be used to study the expected polarisation characteristics for various Compton-scattering based emission models for relativistic jet sources. The code is capable to show how the polarisation changes as a function of the viewing angle and energy of the Compton emission, which can serve as a powerful diagnostic for the  radiation mechanism responsible for e.g. the high-energy emission from blazar jets and GRB prompt emission. The first application of the code will be presented in a follow up paper (Dreyer and B{\"o}ttcher 2020, in preparation) to simulate the polarisation signatures from a model where the BBB in blazar spectra arises from a bulk Compton feature, as proposed by \cite{Baring_etal2017}. The thermal Comptonisation process should lead to significant polarisation of the emission from the UV/$X$-ray excess in the SED. This will reinforce future prospects of using measurements of polarisation signatures to distinguish between different radiation mechanism models for the sources of interest. 
\acknowledgments
We thank the anonymous referee for giving us an expeditious, helpful, and constructive report. The work of M.B. is supported through the South African Research Chair Initiative of the National Research Foundation\footnote{Any opinion, finding and conclusion or recommendation expressed in this material is that of the authors and the NRF does not accept any liability in this regard.} and the Department of Science and Innovation of South Africa, under SARChI Chair grant No. 64789.
\appendix
\section{Comparison to previously published results}\label{sec:MSC_Compare}
The Monte-Carlo code developed by \cite{Krawczynski_2012} can be used to numerically compute the polarisation due to Compton scattering in the Thomson and Klein-Nishina regimes. The numerical results were compared to analytical results of \cite{Bonometto_etal1970} that were based on quantum mechanical scattering calculations in the Thomson regime. The numerical formulation was subsequently used to study the polarisation of Compton radiation emitted in the Klein-Nishina regime. In this Appendix, the results for Compton polarisation in the Thomson and Klein-Nishina regimes from the simulations of the MAPPIES code are compared to the numerical results of \cite{Krawczynski_2012}.

The simulations of \cite{Krawczynski_2012} were in good agreement with the analytical calculations of \cite{Bonometto_etal1970} for Compton polarisation in the Thomson regime. An important implication of the calculations of \cite{Bonometto_etal1970} is that the PD vanishes for unpolarised photons scattered by electrons with Lorentz factors $\gamma \gtrsim 10$. \cite{Krawczynski_2012} tested this prediction by simulating the Compton scattering of an isotropic distribution of $\sim 2.0$ keV mono-energetic, unpolarised seed photons. They found a net PD $\sim 0.26 \%$ due to scattering off mono-energetic electrons with Lorentz factors of $\gamma = 10^3$, assumed to be isotropic in the co-moving frame of the emission region (which moves along the jet with a bulk Lorentz factor of $\Gamma_{jet} = 5$). The MAPPIES code is used to simulate Comptonisation with the same initial conditions to those of \cite{Krawczynski_2012}. The Stokes vectors correspond to a net $\mathrm{PD} \sim 0.09 \%$, and exhibit mean values of 0 when divided into 100 subsets of the simulated events, which is consistent with unpolarised radiation. The results are therefore in good agreement with the results of \cite{Krawczynski_2012}. 

The polarisation for Compton emission in the Klein-Nishina regime was evaluated by \cite{Krawczynski_2012} with simulations of Compton scattering off an isotropic distribution of mono-energetic electrons with Lorentz factors between 10 and 62500. They considered $\sim 1.3$ keV mono-energetic photons, assumed to be fully polarised (with initial Stokes vectors of (I, Q, U) = (1, 1, 0)), and uni-directional in the laboratory frame. A similar setup is used to test the MAPPIES code for Compton polarisation in the Klein-Nishina regime. The seed photons are drawn in the laboratory frame, assumed to be fully polarised, mono-energetic with $kT_{rad} = 3.1$ keV, and unidirectional with $(\Theta_{lab}, \Phi_{lab}) = (1.4, 0)$~rad. 

\begin{figure*}[ht!]
\plotone{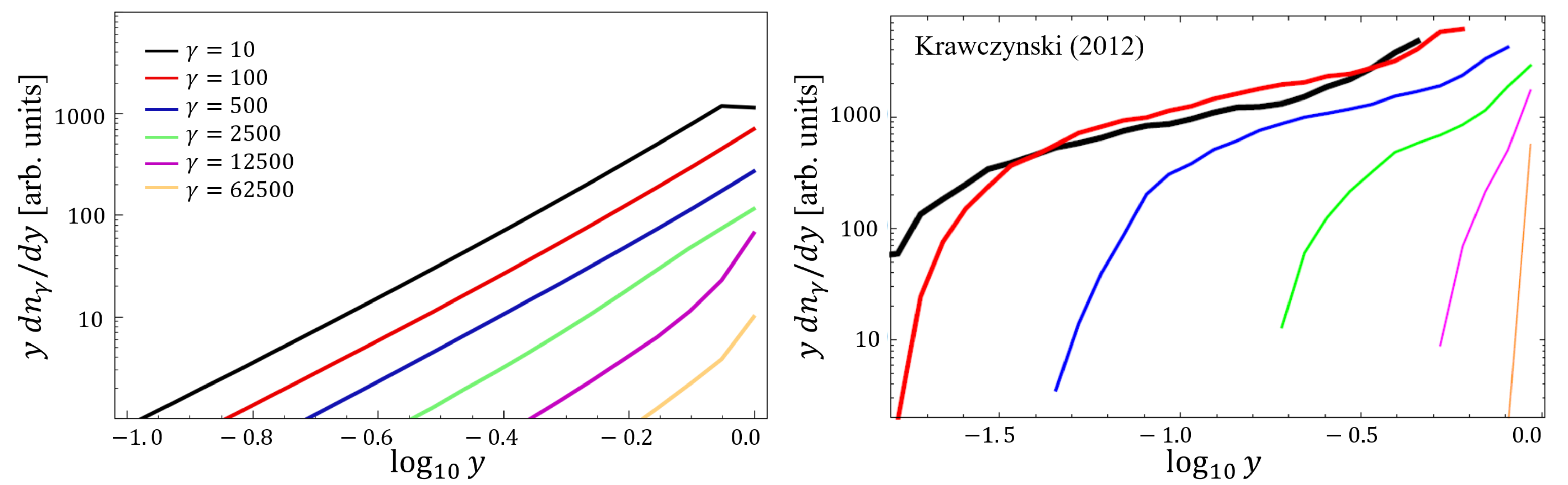}
\caption{The intensity of the Compton emission as a function of the scattered photon energy in units of the maximum kinematically allowed energy $y = x_{lab}^{sc}/x_{lab}^{max}$. The seed photons are uni-directional and mono-energetic in the laboratory frame with $x_{lab} = \epsilon_{lab}/(m_ec^2) = 0.0025$. The results are shown for scattering off mono-energetic electrons, assumed to be isotropic in the co-moving frame of the emission region, with Lorentz factors of $\gamma = 10, 100, 500, 2500, 12500, 62500$. Results from \cite{Krawczynski_2012} are shown in the right panel, and the results from the MAPPIES code are shown in the left panel. \label{fig:app_CompareSpectrum}}
\end{figure*}

\begin{figure*}[ht!]
\plotone{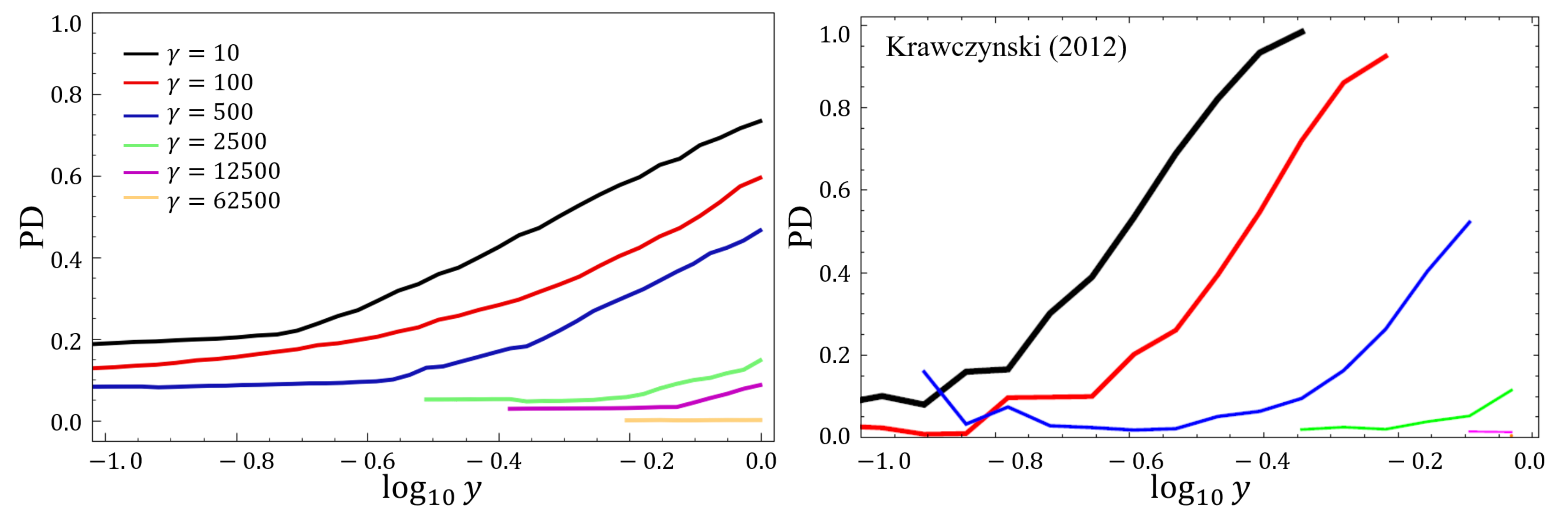}
\caption{The Compton polarisation as a function of the scattered photon energy in units of the maximum kinematically allowed energy $y = x_{lab}^{sc}/x_{lab}^{max}$. The seed photons are uni-directional and mono-energetic in the laboratory frame with $x_{lab} = \epsilon_{lab}/(m_ec^2) = 0.0025$. The results are shown for scattering off mono-energetic electrons, assumed to be isotropic in the co-moving frame of the emission region, with Lorentz factors of $\gamma = 10, 100, 500, 2500, 12500, 62500$. Results from \cite{Krawczynski_2012} are shown in the right panel, and the results from the MAPPIES code are shown in the left panel. \label{fig:app_ComparePD}}
\end{figure*}

\begin{figure*}[ht!]
\plotone{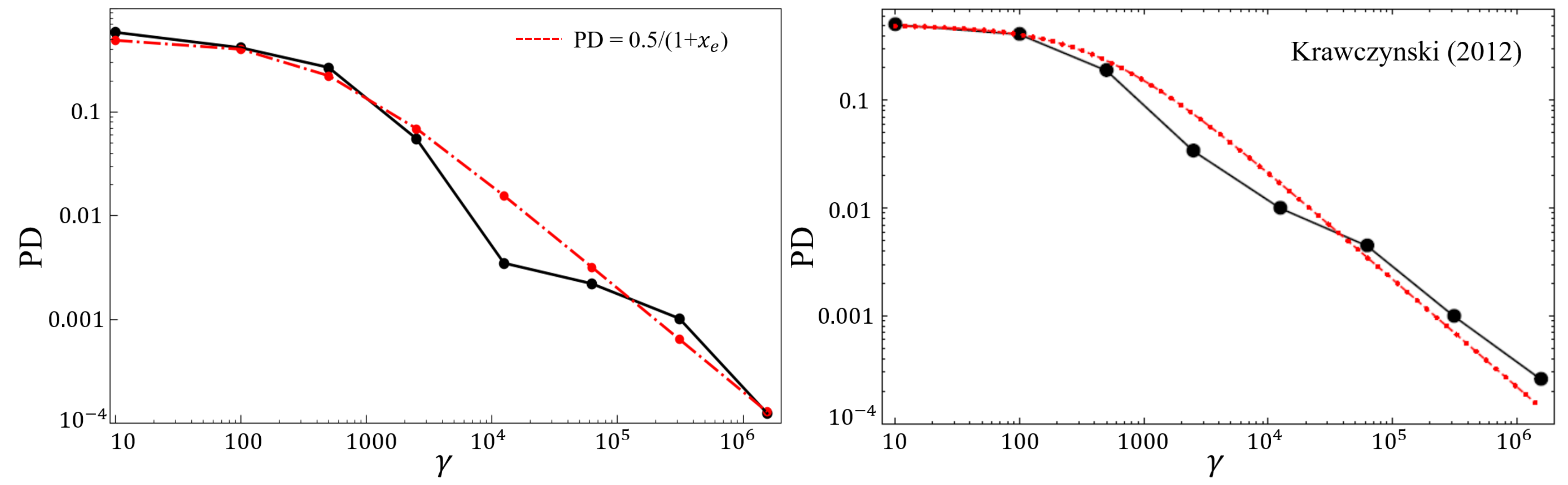}
\caption{The net $\mathrm{PD}$ of the Compton emission as a function of Lorentz factors of the electrons. The seed photons are uni-directional and mono-energetic in the laboratory frame with $x_{lab} = \epsilon_{lab}/(m_ec^2) = 0.0025$. The results are shown for scattering off mono-energetic electrons, assumed to be isotropic in the co-moving frame of the emission region, with Lorentz factors of $\gamma = 10, 100, 500, 2500, 12500, 62500, 3.1 \times 10^5, 1.6\times10^6$. Results from \cite{Krawczynski_2012} are shown in the right panel, and the results from the MAPPIES code are shown in the left panel. The function $\mathrm{PD} = 0.5/(1 + x_e)$ from \cite{Bonometto_etal1970} is indicated with a red line. \label{fig:app_ComparePDvsgamma}}
\end{figure*}

In all figures discussed, the results from the MAPPIES code are shown in the left panels, while the numerical results of \cite{Krawczynski_2012} are shown in the right panels. The intensity and PD are given as a function of the scattered photon energy in Figures \ref{fig:app_CompareSpectrum} and \ref{fig:app_ComparePD}, respectively. The scattered photon energy is shown in units of the maximum energy allowed kinematically $y = \frac{x_{lab}^{sc}}{x_{lab}^{max}}$ where 
\begin{eqnarray}
\label{eq:xMAX}
x_{lab}^{max} = \frac{4\gamma x_{lab}}{1 + 4\gamma x_{lab}}
\end{eqnarray}
with $x_{lab} = \epsilon_{lab}/(m_ec^2)$ and $x_{lab}^{sc} = \epsilon_{lab}^{sc}/(m_ec^2)$ the dimensionless energy of the seed and scattered photons, respectively. The intensity of the scattered photons shifts to higher energies for larger Lorentz factors, and peak towards $x_{lab}^{max}$ deeper in the Klein-Nishina regime. The PD decreases for larger Lorentz factors, and is strongly suppressed for $\gamma \gg 10$. In Figure \ref{fig:app_ComparePDvsgamma}, the net polarisation is shown as a function of the Lorentz factors of the electrons, where the function $\mathrm{PD} = 0.5/(1 + x_e)$ from \cite{Bonometto_etal1970} is indicated with a red line.  The net polarisation decreases approximately with the inverse of the seed photon energy in the electron rest frame, consistent with the analytical prediction of $\mathrm{PD} = 0.5(1 + x_e)$ from \cite{Bonometto_etal1970} in the Thomson regime. The results from the MAPPIES code are therefore overall consistent with those of \cite{Krawczynski_2012}.

\bibliography{REF}{}
\bibliographystyle{aasjournal}
\end{document}